\documentclass[aps,eqsecnum,nofootinbib,showpacs]{revtex4}
\usepackage[dvips]{graphicx}  
\usepackage{amsmath,amsfonts}
\usepackage{bm}
\usepackage{color}  
\newcommand{\ket}[1]{\left|~#1~\right\rangle}

\begin{document}
\title{New Mixing Angles in the Left-Right Symmetric Model}
\author{Akira Kokado}
\email{kokado@kobe-kiu.ac.jp}
\affiliation{Department of Physical Therapy, Kobe International University, Kobe 658-0032, Japan}
\author{Takesi Saito}
\email{tsaito@k7.dion.ne.jp}
\affiliation{Department of Physics, Kwansei Gakuin University,
Sanda 669-1337, Japan}
\date{\today}
\begin{abstract}
 In the left-right symmetric model neutral gauge fields are characterized by three mixing angles $\theta _{12},  \theta_{23}, \theta_{13}$ between three gauge fields 
 $B_\mu , W^3_{L\mu }, W^3_{R\mu }$, which produce mass eigenstates $A_{\mu }, Z_{\mu }, Z'_{\mu }$,  
when $G = SU(2)_L\times SU(2)_R\times U(1)_{B-L} \times D$  is spontaneously broken down until $U(1)_{em}$. 
We find a new mixing angle $\theta '$, which corresponds to the Weinberg angle $\theta _{W}$  in the standard model with the $SU(2)_{L}\times U(1)_{Y}$ gauge symmetry, from these mixing angles. It is then shown that any mixing angle $\theta _{ij}$ can be expressed by $\varepsilon $ and $\theta '$, where  $\varepsilon = g_L/g_R$ is a ratio of running left-right gauge coupling strengths. We observe that light gauge bosons are described by $\theta'$ only, whereas heavy gauge bosons  are described by two parameters $\varepsilon$ and $\theta '$.
\end{abstract}

\maketitle
\section{Introduction}\label{sec:intro}
%
%
Since the discovery of neutrino oscillations \cite{ref:Totsuka}, many new physics beyond the standard model have been proposed in an accelerative way. 
Among these works the left-right symmetric model
(LRSM)\cite{ref:Mohapatra,ref:Duka, ref:Deppisch, ref:Patra, ref:Dev, ref:Borah, ref:Chay} appears to be one of the most promising models, 
which is invariant under the gauge group $G = SU(2)_L\times SU(2)_R\times U(1)_{B-L} \times D$, where $D$ denotes D-parity symmetry 
and $g_L=g_R$ for left-right gauge strengths. The smallness of neutrino masses is one of the most important problems to be solved. 
The so-called seesaw mechanism
\cite{ref:Minkowski, ref:Yanagida, ref:Gell-Mann}  is naturally incorporated in the model, where neutrinos are regarded as Majorana particles. The original gauge group $G$ is first broken down to 
 $SU(2)_L\times SU(2)_R\times U(1)_{B-L}$ at an energy scale $\langle~\eta ~\rangle=M_P\sim 10^{15}$~GeV, where $\eta $ is the pseudoscalar Higgs field, so that D-parity invariance is broken because of the Higgs potential including $\triangle _L$ and $\triangle _R$. Then at the energy scale $\langle~\triangle _R~\rangle=V_R\sim 10^3$~GeV the $SU(2)_R$ invariance is broken until $SU(2)_L\times U(1)_{Y}$, and finally at the energy scale$\langle~\phi ~\rangle=\kappa _{1,2}\sim 10^2$~GeV the last symmetry is broken until the $U(1)_{em}$ invariance. As a result of running couplings we have $g_L \neq g_R$ \cite{ref:Mohapatra}. At this point there are some papers with $g_L = g_R$ in any energy scale \cite{ref:Masso}, whereas our paper is not so.\\
\indent However, except for ambiguity associated with Higgs field, some unsolved problems
remain in various gauge fields. 
Light gauge bosons $W^{\pm}$ and $Z$ are associated with their partners $W'^{\pm}$ and $Z'$ which
are too heavy to observe. For charged gauge bosons $W^{\pm}$ and $W'^{\pm}$ they are
characterized by one mixing angle $\gamma $. On the other hand, for neutral gauge fields they
are characterized by three mixing angles $\theta _{12}, \theta_{23}, \theta_{13}$ 
between three gauge fields $B_\mu , W^3_{L\mu }, W^3_{R\mu },$ which produce mass eigenstates $A_{\mu }, Z_{\mu } , Z'_{\mu}$. 
In addition there are free gauge parameters $g_L, g_R$ and $g_1$ of $G$. \\
\indent Now we are faced with problems how to fix these parameters from experimental data presently available such as $W^{\pm}, Z$. Apparently some of parameters remain unfixed, until when we have data from their heavy partners $W'^{\pm}, Z'$. \\
\indent In this paper we discuss these problems. The mass matrix for neutral gauge fields will be diagonal if $\tan{\theta _{23}}=-\sin{\theta _{12}}\sin{\theta_{13}}/\cos{\theta_{12}} + O(\delta )$ , where $\delta $ is an infinitesimally small parameter associated with the spontaneously broken left-right symmetry. In $\delta \sim  0$, then introducing a new mixing angle $\theta '$ defined by  $s'=\sin{\theta _{12}}\cos{\theta _{13}}$ and $c'=\cos{\theta _{12}}/\cos{\theta _{23}}$ with $s'^2 +  c'^2 =1$ ( $s' , c'$ stand for
$\sin{\theta '}, \cos{\theta'}$, respectively), we show that all light gauge boson masses can be 
 expressed in terms of $s', c'$, namely, $M_W = 37.3/s'$~GeV and $M_Z/M_W = 1/c'$. The coupling
strength between the proton and the $Z$ boson as well as those of neutrinos and $Z$
can be shown to be expressed also by $s', c'$. All results are completely the same as those of the Weinberg-Salam (WS) theory with $SU(2)_{L}\times U(1)_{Y}$, in $\delta \sim 0$, if we take  $s'= \sin{\theta_{W} }$, $c'=\cos{\theta_{W}}$, where $\theta _W$ is the Weinberg angle. Here WS gauge coupling constants $g, g'$ are given by $g=g_L=e_0/\sin{\theta _W}$ and $g'=g_1\cos{\theta _{13}}\cos{\theta _{23}}=e_0/\cos{\theta _W}$  with $e_0$ the positron charge, respectively. These results happen to be around mass scales of weak gauge bosons $M_W, M_Z$.\\
\indent We will also show that any mixing angle $\theta _{ij}$ can be expressed in terms of $\theta '=\theta _{W}$ and $\varepsilon = g_L/g_R$ in $\delta \sim  0$.\\
\indent Finally we discuss about the gauge coupling ratio $\varepsilon$, which is given by angles as $\varepsilon ^2 = s'^{-2} - \sin{\theta _{12}}^{-2}$. Hence we cannot fix $\varepsilon $ by $s'$ only. We then observe generally that light gauge bosons are described by $\theta '$ only, whereas heavy gauge bosons are described by two parameters $\varepsilon $ and $\theta '$. \\ 
\indent In Secs.\ref{sec:2} and \ref{sec:3} we summarize the LRSM in order to fix the notation.  
In Secs.\ref{sec:4} and \ref{sec:5} masses of gauge bosons are calculated in the order of $\delta $. In Secs.\ref{sec:6} we discuss mixing angles. In the Appendix we calculate the electromagnetic coupling strength with fermions, and also the coupling strength between the proton and the $Z$ boson. The final section is devoted to concluding remarks.
%
\section{The Left-Right Symmetric Model}\label{sec:2}
Let us summarize the LRSM proposed first in Ref. \cite{ref:Mohapatra}, which is invariant under the
gauge group 
\begin{align}
 &  G =  SU(2)_L \times SU(2)_R \times U(1)_{B-L} \times D~.
  \label{eq:Group}
\end{align}
The representation of $G$ is characterized by the triplet ($d_L , d_R ,Y$), where $d_L, d_R$ 
denote the dimensions of $SU(2)_L$ and $SU(2)_R$, respectively, and $Y$ is defined by $Q=I_{3L} + I_{3R} +Y/2$  in familiar notations and is equal to
$Y=B-L$. The fermion doublets are given by
\begin{align}
 &  \mbox{quarks:} \quad Q_L(2, 1, \frac{1}{3}), \quad Q_R(1, 2, \frac{1}{3})~,
  \label{eq:fermion_doublets} \\
 &  \mbox{leptons:} \quad L_L(2, 1, -1), \quad L_R(1, 2, -1)~.
  \nonumber
\end{align}
We also introduce four kinds of Higgs fields 
\begin{align}
 & \phi (2, 2, 0)~, \quad \Delta _L(3, 1, 2)~, \quad \Delta _R(1, 3, 2)~, \quad \eta (1, 1, 0)~,
  \label{eq:Hidggs_fields}
\end{align}
with representations as
\begin{align}
 & \phi  = \left(
   \begin{array}{cc}
      \phi _1^0 & \phi _1^{+} \\
      \phi_2^- & \phi_2^0 \\
   \end{array}
  \right)~,
 \quad
 \tilde {\phi } = \left(
   \begin{array}{cc}
      \bar {\phi }_2^0 & -\phi _2^{+} \\
      -\phi_1^- & \bar{\phi}_1^0 \\
   \end{array}
  \right)~,
 \label{eq:matrix_phi} \\
 & \Delta _L = \left(
   \begin{array}{cc}
      \delta _L^+ /\sqrt{2} & \delta _L^{++} \\
      \delta _L^0 & -\delta _L^+ /\sqrt{2} \\
   \end{array}
  \right)~, 
  \quad
 \Delta _R = \left(
   \begin{array}{cc}
      \delta _R^+ /\sqrt{2} & \delta _R^{++} \\
      \delta _R^0 & -\delta _R^+ /\sqrt{2} \\
   \end{array}
  \right)~.
\nonumber
\end{align}
\indent We consider only leptonic parts :
\begin{align}
 & L = L_F + L_Y  + L_B~.
 \label{eq:Action_leptonic_part} 
\end{align}
with
\begin{align}
 & L_F = i \bar{\psi}_L^{j} \gamma ^\mu \Big(\partial _\mu - i\frac{1}{2} g_1 Y_F B_\mu - i g_L W_\mu ^L \Big) \psi _L^j~
 \label{eq:Action_fermion_part} \\
& + i \bar{\psi}_R^{j} \gamma ^\mu \Big(\partial _\mu - i\frac{1}{2} g_1 Y_F B_\mu - i g_R W_\mu ^R \Big) \psi _R^j~, \quad W_{\mu }^{L, R} \equiv \frac{1}{2}\tau _{\alpha }W_{\mu ,L ,R}^\alpha ~,
 \nonumber  
\end{align}
\begin{align}
 & L_Y  = -\bar{\psi}_L^{i} \Big(f_{ij}\phi + \tilde {f}_{ij}\tilde{\phi }  \Big) \psi _R^j~ +  \mbox{H. c.} 
 \label{eq:Action_gamma_part} \\
 & \quad - i \psi_L ^{Ti} C h^{L}_{ij} \tau _{2} \Delta _{L}\psi _L^j + \mbox{H. c.}
 \nonumber  \\
 & \quad - i \psi_R ^{Ti} C h^{R}_{ij} \tau _{2} \Delta _{R}\psi _R^j + \mbox{H. c.}~,
 \nonumber
\end{align}
\begin{align}
 & L_B = tr|D_\mu \Delta _L|^2 + tr|D_\mu \Delta _R|^2 + tr|D_\mu \phi |^2 + \frac{1}{2} \partial _\mu \eta \partial ^\mu \eta 
 \label{eq:Action_Boson_part} \\
 & \quad + \mbox{Yang-Mills terms of } B_{\mu }, W_\mu ^L, W_\mu ^R
 \nonumber \\
 & \quad - V(\mbox{Higgs potential of } \phi , \Delta _L, \Delta _R, \eta )~,
 \nonumber
\end{align}
\begin{align}
 & \psi _L^i  = \left(
   \begin{array}{c}
      \nu _i \\
      e_i \\
   \end{array}
  \right)_L~,
  \quad 
  \psi _R^i  = \left(
   \begin{array}{c}
      \nu _i \\
      e_i \\
   \end{array}
  \right)_R
  \quad
  i = e, \mu , \tau , 
  \label{eq:fermion_lepton} \\
 & \tilde {\phi } = \tau _2 \phi ^* \tau _2 ~.
 \nonumber
\end{align}
The Dirac-Majorana type couplings are given in Eq.(\ref{eq:Action_gamma_part}). \\
\indent Under gauge transformations Higgs fields are transformed as
\begin{align}
 & \phi \to U_L \phi U_R^{-1}~, \quad \tilde {\phi} \to U_L \tilde{\phi} U_R^{-1}~,
 \label{eq:Unitrary_tr_phi} \\
 & \Delta _L \to U_L \Delta_L U_L^{-1}~, \quad \Delta _R \to U_R \Delta_R U_R^{-1}~,
 \nonumber 
\end{align}
so that covariant derivatives are given by
\begin{align}
 & D\phi  = \partial \phi  - i \big( g_L W_L \phi  - g_R \phi W_R \big)~,
 \label{eq:deribative_phi} \\
 & D\Delta _L = \partial \Delta _L - i \frac{1}{2}g_1 Y_\Delta B \Delta _L - i g_L (W_L \Delta _L - \Delta _L W_L ) ~,
 \nonumber \\
 & D\Delta _R = \partial \Delta _R - i \frac{1}{2}g_1 Y_\Delta B \Delta _R - i g_R (W_R \Delta _R - \Delta _R W_R ) ~,
  \nonumber
\end{align}
where $Y_\phi = 0, Y_\Delta = +2$ and $Y_F = -1$. \\
\indent The gauge symmetry of $SU(2)_L\times  SU(2)_R\times U(1)_{B-L} \times D$ is broken down until $U(1)_{em}$ by nonzero vacuum expectation values of Higgs fields
\begin{align}
 & \left\langle~\phi~\right\rangle  = \frac{1}{\sqrt{2}}\left(
   \begin{array}{cc}
      \kappa _1 & 0 \\
      0 & \kappa _2 \\
   \end{array}
  \right)~,
 \quad
 \left\langle~\Delta _{L,R}~\right\rangle  = \frac{1}{\sqrt{2}}\left(
   \begin{array}{cc}
      0 & 0 \\
      V_{L,R} & 0 \\
   \end{array}
  \right)~,
  \quad
  \left\langle~\eta ~\right\rangle = \frac{1}{\sqrt{2}}\eta _0 ~,
 \label{eq:VEV}
\end{align}
with assumptions
\begin{align}
 & \big| V_L \big| << \big| \kappa _{1, 2} \big| <<  \big|V_R \big| << \big| \eta _0 \big|~.
 \label{eq:inequality1}
\end{align}
\indent The Yukawa terms then become
\begin{align}
 & -L_Y  = \bar{\psi}_L^{i} \Big(f_{ij}\left\langle\phi \right\rangle + \tilde {f}_{ij}\left\langle\tilde{\phi }\right\rangle \Big) \psi _R^j~ +  \mbox{H. c.} 
 \label{eq:VEV_gamma_part} \\
 & \quad + i \psi_L ^{Ti} C \tau _{2} h^{L}_{ij} \left\langle\Delta _{L}\right\rangle\psi _L^j + \mbox{H. c.}
 \nonumber  \\
 & \quad + i \psi_R ^{Ti} C \tau _{2} h^{R}_{ij} \left\langle\Delta _{R}\right\rangle\psi _R^j + \mbox{H. c.}
 \nonumber \\
 & = \bar {\nu }_L^i m_{Dij} \nu _R^{j} + \frac{1}{2} \bar {\nu }_R^{iC} m_{Lij} \nu _L^{j}
  + \frac{1}{2} \bar {\nu }_L^{iC} m_{Rij} \nu _R^{j} + \bar{e}_L^i m'_{Dij} e_R^j + \mbox{H. c.}~.
  \nonumber
\end{align}
Here we have used notations
\begin{align}
 & \bar{\psi }^C_L =\big( \frac{1-\gamma _5}{2}\psi ^C\big)^{\dagger } \gamma ^0= \psi ^T_R C~, \quad \bar{\psi }^C_R  = \psi ^T_L C ~,
 \label{eq:notation_C}
\end{align}
and
\begin{align}
 & m_{Dij} = \frac{f_{ij} \kappa _1 + \tilde {f}_{ij} \kappa _2 }{\sqrt{2}}\simeq m_l~,
 \quad m_{Rij} = \sqrt{2} V_R h^R_{ij}~, \quad m_{Lij} = \sqrt{2} V_L h_{ij}^L ~,
 \label{eq:mass1} \\
 & m'_{Dij} = \frac{f_{ij} \kappa _2 + \tilde {f}_{ij} \kappa _1}{\sqrt{2}}\simeq m_l~.
 \nonumber
\end{align}
In order to make mass terms real, any parameter here is taken to be real. We assume
that $m_D$ and $m'_D$ take values of the order of leptonic masses $m_l$, i.e., $m_D \simeq m'_D \simeq  m_l$~. \\
\indent According to Eq.(\ref{eq:inequality1}) we have inequalities
\begin{align}
 & \big| m_{Lij} \big| << \big| m_{Dij} \big| <<  \big|m_{Rij} \big|~.
 \label{eq:inequality2}
\end{align}
The neutrino terms can be written as
\begin{align}
 & \left(
   \begin{array}{cc}
      \bar {\nu } & \bar{N} \\
   \end{array}
  \right)
   \left(
   \begin{array}{cc}
      m _L & m_D \\
      m_D & m _R \\
   \end{array}
  \right)
  \left(
   \begin{array}{c}
      \nu  \\
      N \\
   \end{array}
  \right)
   = \bar{\nu } m_L \nu + \bar{\nu} m_D N + \bar{N} m_D \nu + \bar{N} m_R N ~,
 \label{eq:neutrino_mass_term} 
\end{align}
where
\begin{align}
 & \nu = \frac{\nu _L + (\nu_L)^C}{\sqrt{2}} = \frac{\nu _L + \nu _R^C}{\sqrt{2}}~,
 \label{eq:lepton_LR1} \\
 & N = \frac{\nu _R + (\nu_R)^C}{\sqrt{2}} = \frac{\nu _R + \nu _L^C}{\sqrt{2}}~.
  \label{eq:lepton_LR2}
\end{align}
Making the mass matrix diagonal we have the seesaw result for each $ij$
\begin{align}
 & \big| m_{aij}\big| \simeq \big|\frac{(m_{Dij})^2}{m_{Rij}}\big|~,
 \label{eq:seesaw_result1} \\
 & \big| m_b \big| \simeq \big| m_{Rij} \big|~,
 \label{eq:seesaw_result2}
\end{align}
with eigenstates
\begin{align}
 & \ket{\nu _{aj}} \simeq \ket{\nu _{j}} - \frac{m_{Dij}}{m_{Rij}} \ket{N_j}~,
 \label{eq:ket_equation1} \\
 & \ket{N_{bj}} \simeq \ket{N_{j}} + \frac{m_{Dij}}{m_{Rij}} \ket{\nu _j}~.
 \label{eq:ket_equation2} 
\end{align}
Finally Eq.(\ref{eq:neutrino_mass_term}) turns out to be
\begin{align}
 & \bar{\nu }_{ai} m_{aij} \nu _{aj} + \bar{N}_{bj} m_{bij} N_{bj}~.
 \label{eq:neutrion_mass_term2}
\end{align}
Each term should be diagonal for three generations, $i, j = 1,2,3$. \\
\indent The $\nu $ is made of $\nu _L$, while $N$ is made of $\nu _R$, as is seen from Eqs.(\ref{eq:lepton_LR1}) 
and(\ref{eq:lepton_LR2}). The mass of $\nu $ is so small, whereas that of $N$ is too heavy to observe,
according to the seesaw mechanism. This is the reason why $\nu _L$  only is observable but
not $\nu _R$. The heavy particle $\nu _R$ together with $W'$ and $Z'$ is now widely recognized as
one of strong candidates for dark matters.
%
\section{The $T$ matrix} \label{sec:3}
Let us introduce new variables $(A_\mu , Z_\mu , Z'_\mu )$, which are mass eigenstates of three gauge fields 
$(B_\mu , W^3_{L\mu }, W^3_{R\mu })$, by 
\begin{align}
 &   \left(
   \begin{array}{c}
      B \\
      W_L^3 \\
      W_R^3 \\
   \end{array}
  \right)
   = T
	\left(
   \begin{array}{c}
      A \\
      Z \\
      Z' \\
   \end{array}
  \right)~,
 \label{eq:T_matrix} 
\end{align}
where $T$ is a $3\times 3$ unitary matrix. Then we have
\begin{align}
 & B = T_{11} A + T_{12} Z + T_{13} Z'~,
 \label{eq:T_matrix2} \\
 & W_L^3 = T_{21} A + T_{22} Z + T_{23} Z'~,
 \nonumber \\
 & W_R^3 = T_{31} A + T_{32} Z + T_{33} Z'~.
 \nonumber 
\end{align}
From gauge coupling terms in Lagrangian (\ref{eq:Action_fermion_part}) a collection of coefficients of $A_\mu (x)$ suggest us that the electric charge $-e_0$ of the electron is given by, as shown in Appendix,
\begin{align}
 & e_0 = \frac{1}{2}( g_1 T_{11} + g_L T_{21})~,
 \label{eq:coupling_const} \\
 & g_1 T_{11} = g_{L} T_{21} = g_R T_{31} = e_0~.
 \nonumber
\end{align}
Here we have the desired electromagnetic coupling term $e_0 A \bar{e}\gamma e$. \\
\indent The $T$ matrix is generally given by three mixing angles
\begin{align}
 &  T = \left(
   \begin{array}{ccc}
      c_{12} & -s_{12} & 0 \\
      s_{12} & c_{12}  & 0 \\
      0      &   0     & 1 \\
   \end{array}
  \right)
  \left(
   \begin{array}{ccc}
      c_{13} & 0  & -s_{13} \\
      0 & 1  & 0 \\
      s_{13}      &   0     & c_{13} \\
   \end{array}
  \right)
    \left(
   \begin{array}{ccc}
      1  & 0  & 0 \\
      0  & c_{23}  & -s_{23} \\
      0  &  s_{23}     & c_{23} \\
   \end{array}
  \right)
 \label{eq:T_matrix3} \\
 & = \left(
   \begin{array}{ccc}
      c_{12}c_{13} & -s_{12}c_{23}-c_{12}s_{23}s_{13} & s_{12}s_{23} - c_{12}c_{23}s_{13} \\
      s_{12}c_{13} & c_{12}c_{23} - s_{12}s_{23}s_{13}  & -c_{12}s_{23}-s_{12}c_{23}s_{13} \\
      s_{13}     &   s_{23}c_{13}     & c_{23}c_{13} \\
   \end{array}
  \right)~,
\end{align}
where $c_{ij}=\cos{\theta _{ij}}, s_{ij}=\sin{\theta_{ij}}$, and sometimes we use a notation $\theta_{12}= \theta $ for simplicity. From constraints (\ref{eq:coupling_const}) we have
\begin{align}
 & g_1 T_{11} = g_1c_{12}c_{13} = e_0~, 
 \label{eq:coupling_const_relation} \\
 & g_L T_{21} = g_L s_{12}c_{13} = e_0~,
 \nonumber \\
 & g_R T_{31} = g_R s_{13} = e_0~, 
  \nonumber
\end{align}
so that 
\begin{align}
 & \tan{\theta _{12}} = \frac{g_1}{g_L}~, 
 \label{eq:coupling_const_relation2} \\
 & \sin{\theta _{13}} = \frac{e_0}{g_R}~,
 \nonumber \\
 & \frac{1}{e_o^2} = \frac{1}{g_1^2} + \frac{1}{g_L^2} + \frac{1}{g_R^2}~,
 \nonumber \\
 & g_1 \cos{\theta } = \frac{e_o}{c_{13}}~, \quad g_L \sin{\theta } = \frac{e_o}{c_{13}}~, \quad \theta = \theta _{12}~. 
 \nonumber
\end{align}
%
%
\section{Masses of charged gauge bosons} \label{sec:4}
The aim of this section is to calculate the numerical values of $\delta $ and $V_{R}$, which are defined in Eq.(\ref{eq:ratio_MW_MW'}). In order to do this, we follow the work of Ref.\cite{ref:Langacker}.\\
\indent After the gauge symmetry is spontaneously broken, the trace parts in Eq.(\ref{eq:Action_Boson_part}) are summarized as follows:
\begin{align}
 & tr|D \Delta _L|^2 = \frac{1}{2}\big[|g_LW_L^+|^2 + (g_1B - g_LW_L^3)^2 \big] V_L^2~,
 \label{eq:diff_delta_L} \\
 & tr|D \Delta _L|^2 = \frac{1}{2}\big[|g_RW_R^+|^2 + (g_1B - g_RW_R^3)^2 \big] V_R^2~,
 \nonumber \\
 & tr|D \phi |^2 = \frac{1}{8}\big[(g_LW_L^3 - g_RW_R^3)^2(\kappa _1^2 + \kappa _2^2) + 2|\kappa _1g_LW_L^+ - \kappa _2g_RW_R^+|^2 + 2|\kappa _2 g_LW_L^+ - \kappa _1 g_RW_R^+|^2 \big]~,
 \nonumber \\
 & W_{L, R}^{\pm} = (W^1 \mp iW^2)_{L, R}/\sqrt{2}~.
 \nonumber
\end{align}
\indent The charged boson mass terms are then given by
\begin{align}
 &  X \equiv \frac{1}{2} |g_L W_L^+|^2 V_L^2 + \frac{1}{2} |g_R W_R^+|^2 V_R^2
 \label{eq:def_boson_mass} \\
 & + \frac{1}{4} |\kappa _1 g_L W_L^+ - \kappa _2 g_R W_R^+ |^2 + \frac{1}{4} |\kappa _2 g_L W_L^+ - \kappa _1 g_R W_R^+ |^2
  \nonumber \\
 & = \frac{1}{2} |W_L^+|^2 \big[g_L^2 V_L^2 + \frac{1}{2}(\kappa _1^2 + \kappa _2^2 ) g_L^2\big] + \frac{1}{2}|W_R^+|^2 \big[g_R^2 V_R^2 + \frac{1}{2}(\kappa _1^2 + \kappa _2^2 ) g_R^2\big]
  \nonumber \\
 & - \frac{1}{2} \kappa _1 \kappa _2 g_L g_R (W_L^+ W_R^- + W_R^+ W_L^-)~.
 \nonumber
\end{align}
\indent Let us introduce here new variables as mass eigenstates by 
\begin{align}
 &   \left(
   \begin{array}{c}
      W_L^\pm \\
      W_R^\pm \\
   \end{array}
  \right)
   = U
	\left(
   \begin{array}{c}
      W^\pm \\
      W'^\pm \\
   \end{array}
  \right)~,
  \quad
  U =
  	\left(
   \begin{array}{cc}
      \cos{\gamma } & \sin{\gamma } \\
     -\sin{\gamma } & \cos{\gamma } \\
   \end{array}
  \right)~.
 \label{eq:U_matrix} 
\end{align}
Eq.(\ref{eq:def_boson_mass}) turns out to be of the form
\begin{align}
 & X=|W|^2 M_W^2 + |W'|^2 M_{W'}^2 + (W^{\dagger }W' + W'^{\dagger }W)\lambda ,
 \label{eq:def_boson_mass2}
\end{align}
where
\begin{align}
 & M_W^2 = G_L U_{11}^2 + G_R U_{21}^2 - \kappa _1 \kappa _2 g_L g_R U_{11} U_{21} ~,
 \label{eq:mass_M_W1} \\
 & M_{W'}^2 = G_L U_{12}^2 + G_R U_{22}^2 - \kappa _1 \kappa _2 g_L g_R U_{12} U_{22} ~,
 \label{eq:mass_M_W'1} \\
 & \lambda = U_{11} U_{12} G_L + U_{21} U_{22} G_R - \frac{1}{2} \kappa _1 \kappa _2 g_L g_R (U_{12} U_{21} + U_{22} U_{11})~,
 \label{eq:mass_lambda} 
\end{align}
with
\begin{align}
 & G_L = \frac{1}{2} g_L^2 \big[ V_L^2 + \frac{1}{2}(\kappa _1^2 + \kappa _2^2)]~,
 \quad 
 G_R = \frac{1}{2} g_R^2 \big[ V_R^2 + \frac{1}{2}(\kappa _1^2 + \kappa _2^2)]~.
 \label{eq:u_eq2}
\end{align}
\indent The vanishing condition of the cross term is given by
\begin{align}
 & \lambda = 0~,
 \label{eq:def_lambda}
\end{align}
\begin{align}
 & \mbox{i.e.} \quad \tan{2\gamma } = \frac{\kappa _1 \kappa _2 g_L g_R}{G_L - G_R} \simeq -\frac{2\kappa _1 \kappa _2}{V_R^2}\varepsilon = 2\gamma 
\end{align}
hence
\begin{align}
 & \Big|\frac{\gamma }{\varepsilon }\Big|= \Big|\frac{\kappa _1 \kappa _2}{V_R^2}\Big| << 1~,
 \label{eq:gamma_very_small} 
\end{align}
where $\varepsilon =g_L/g_R$. We see that the mixing angle $\gamma $ is very small. \\
\indent Finally we find the charged boson masses
\begin{align}
 & M_W^2 = g_L^2\frac{\kappa _1^2 + \kappa _2^2}{4} + \frac{1}{2} g_L^2 \kappa _1 \kappa _2\big(\frac{\gamma }{\varepsilon }\big) \simeq g_L^2\frac{\kappa _1^2 + \kappa _2^2}{4}~,
 \label{eq:mass_M_W2} \\
 & M_{W'}^2 = \frac{1}{2} g_R^2 V_R^2 - g_L^2 \kappa _1 \kappa _2\big(\frac{\gamma }{\varepsilon }\big)
\simeq \frac{1}{2} g_R^2 V_R^2~,
\label{eq:mass_M_W'}
\end{align}
Substituting $V_R^2$ and $g_L^2$ into Eq.(4.10), the mixing angle can be written as
\begin{align}
 & \tan{2\gamma } = - \frac{4\kappa _1 \kappa _2}{\kappa _1^2 + \kappa _2^2} \frac{g_R}{g_L}\Big( \frac{M_W}{M_{W'}} \Big)^2~.
 \label{eq:cal_gamma2}
\end{align}
Also the mass ratios of $W$ and $W'$ are given by 
\begin{align}
 & \Big( \frac{M_W}{M_{W'}} \Big)^2 = 2 \Big( \frac{g_L}{g_R}\Big)^2 \frac{\kappa _1^2 + \kappa _2^2}{4V_R^2} = 2 \varepsilon ^2 \delta~, \quad \delta \equiv \frac{\kappa _1^2 + \kappa _2^2}{4V_R^2}~.
 \label{eq:ratio_MW_MW'}
\end{align}
\indent These formulas are completely identical with results of Ref.\cite{ref:Langacker}. \\
If we take values of Deppisch et al. \cite{ref:Deppisch}, $g_L=0.63, g_R=0.38$, and the recent LHC data of $M_{W'}=1900$~GeV\cite{ref:Brehmer}, together with $M_{W}=80$~GeV, then we have $\delta =3.2\times 10^{-4}$. Furthermore, we get the vacuum expectation value $V_R=7.1\times 10^3$~GeV from Eq.(\ref{eq:mass_M_W'}). If we take values of Dev et al. \cite{ref:Dev}, $g_L=0.63, g_R=0.51$, then we have $\delta =5.8\times 10^{-4}$ and $V_R=5.3\times 10^3$~GeV. Finally we get $V_{\phi }=\sqrt{(\kappa _1^2 + \kappa _2^2)/4}\sim $127~GeV from $M_W=g_LV_{\phi}$ .
%
\section{Masses of neutral gauge bosons} \label{sec:5}
%
The aim of this section is to find constraint equations Eq.(\ref{eq:solution_theta23}) among mixing angles, which allow us to introduce a new mixing angle in Sec.\ref{sec:6}. \\
\indent From Eq.(\ref{eq:diff_delta_L}) the mass terms of neutral gauge boson are collected as
\begin{align}
 & X' \equiv \frac{1}{2} \big[ (g_1 B - g_L W_L^3)^2 V_L^2 + (g_1 B - g_R W_R^3)^2 V_R^2 \big] +
 \frac{1}{8} \big[ (g_L W_L^3 - g_R W_R^3)^2(\kappa _1^2 + \kappa _2^2)\big]~,
 \label{eq:mass_neutral_boson} 
\end{align}
where
\begin{align}
 & g_1 B - g_L W_L^3 = Z(g_1T_{12} - g_L T_{22}) + Z'(g_1 T_{13} - g_L T_{23})~,
 \label{eq:part1_mass_neutral_boson} \\
 & g_1 B - g_R W_R^3 = Z(g_1T_{12} - g_R T_{32}) + Z'(g_1 T_{13} - g_R T_{33})~,
 \label{eq:part2_mass_neutral_boson} \\
 & g_L W_L^3 - g_R W_R^3 = Z(g_LT_{22} - g_R T_{32}) + Z'(g_L T_{23} - g_R T_{33})~.
 \label{eq:part3_mass_neutral_boson}
\end{align}
Substituting Eqs.(\ref{eq:part1_mass_neutral_boson})-(\ref{eq:part3_mass_neutral_boson}) into Eq.(\ref{eq:mass_neutral_boson}) we get
\begin{align}
 & X' = \frac{1}{2} Z^2 M_Z^2 + \frac{1}{2} Z'^2 M_{Z'}^2 + \mu  Z Z'~,
 \label{eq:mass_neutral_boson2} 
\end{align}
where
\begin{align}
 & M_Z^2 = (g_1 T_{12} - g_L T_{22})^2 V_L^2 + (g_1 T_{12} - g_R T_{32})^2 V_R^2 + \frac{\kappa _1^2 + \kappa _2^2}{4} (g_L T_{22} - g_R T_{32})^2~,
 \label{eq:mass_M_Z} \\
 & M_{Z'}^2 = (g_1 T_{13} - g_L T_{23})^2 V_L^2 + (g_1 T_{13} - g_R T_{33})^2 V_R^2 + \frac{\kappa _1^2 + \kappa _2^2}{4} (g_L T_{23} - g_R T_{33})^2~,
 \label{eq:mass_M_Z'} \\ 
 & \mu  = (g_1 T_{12} - g_L T_{22})(g_1 T_{13} - g_L T_{23}) V_L^2 + (g_1 T_{12} - g_R T_{32})(g_1 T_{13} - g_R T_{33}) V_R^2 
 \label{eq:mass_nu} \\
 & + \frac{\kappa _1^2 + \kappa _2^2}{4}(g_L T_{22} - g_R T_{32})(g_L T_{23} - g_R T_{33})~.
 \nonumber
\end{align}
The vanishing condition of the cross term is given by
\begin{align}
 & \mu  = 0~.
 \label{eq:nu_vanish}
\end{align}
If we put $V_L=0$,  then Eq.(\ref{eq:nu_vanish}) turns out to be
\begin{align}
 & (g_1 s c_{23} + l s_{23})(g_1 s s_{23} - l c_{23}) = \delta (g_L c c_{23} - l s_{23})(-g_L c s_{23} - l c_{23})~,
  \label{eq:eq_nu}
\end{align}
where
\begin{align}
 & \delta \equiv \frac{\kappa _1^2 + \kappa _2^2}{4 V_R^2}<<1~, \quad l\equiv \sqrt{g_R^2 + g_1^2 c^2}~.
 \nonumber
\end{align}
Solutions to $\theta _{23}$ are
\begin{align}
 & \tan{\theta _{23}} = \frac{l}{g_1 s} + O(\delta )~, \quad -\frac{g_1 s}{l} + O(\delta )~.
 \label{eq:solution_theta23}
\end{align}
\indent The $Z$ boson mass is then given by Eq. (\ref{eq:mass_M_Z}), which reduces to
\begin{align}
 & M_Z^2 = V_R^2(g_1sc_{23} + l s_{23})^2 + \frac{\kappa _1^2 + \kappa _2^2}{4}(g_L c c_{23} - l s_{23})^2 
= \frac{\kappa _1^2 + \kappa _2^2}{4}(g_L^2 + g_1^2) c_{23}^2~, 
 \label{eq:mass_Z_2}
\end{align}
This final form can be seen as follows: Substituting the second solution of Eq.(\ref{eq:solution_theta23}) into the second bracket with $V_R^2$  in Eq.(\ref{eq:mass_M_Z}), we get
\begin{align}
 & V_R^2(g_1sc_{23} + l s_{23})^2 = V_R^2c_{23}^2(g_1 s + l \tan{\theta _{23}})^2 = V_R^2 c_{23}^2 l^2 \times O(\delta ^2) \sim \frac{1}{V_R^2}~.
 \label{eq:first_bracket} 
\end{align}
The third bracket term becomes
\begin{align}
 & g_L c c_{23} - l s_{23} = c_{23} ( g_L c - l \tan{\theta _{23}}) = c_{23} (g_L c + g_1 s -O(l\delta )) = c_{23}\sqrt{g_L^2 + g_1^2} - O(lc_{23}\delta )~.
  \label{eq:second_bracket}
\end{align}
Neglecting $O(\delta )$ term we get the final result of Eq. (\ref{eq:mass_Z_2}). \\
\indent The first solution in Eq.(\ref{eq:solution_theta23}) is inadequate for $M_Z$, because it gives a big mass of order $V_R^2$. \\
\indent In the same way we have

\begin{align}
  & M_{Z'}^2 = c_{23}^2 \big(l + \frac{g_1 s^2}{l}\big)^2V_R^2 + \frac{\kappa _1^2 + \kappa _2^2}{4}c_{23}^2(l - \frac{g_L g_1 sc}{l})^2~,
 \label{eq:mass_M_Z'_2} 
\end{align}
or
\begin{align}
 & M_{Z'} \simeq l c_{23} \big(1 + \frac{g_1 s^2}{l^2}\big)V_R=g_R V_R \frac{1}{c_{23}c_{13}}~. 
 \label{eq:mass_MZ'_3}
\end{align}
%
%
%
%
\section{Mixing angles} \label{sec:6}
%
In this section we consider what kinds of mixing angles are observable. At first we derive a relation
\begin{align}
 & \frac{M_Z}{M_W} = \frac{\sqrt{g_1^2 + g_L^2}}{g_L} c_{23} = \frac{\cos{\theta _{23}}}{\cos{\theta }}~,
 \label{eq:ratio_MZ_MW} 
\end{align}
which follows from Eqs. (\ref{eq:mass_M_W2}) and (\ref{eq:mass_Z_2}). From Eq.(\ref{eq:coupling_const_relation2}) and $l=\sqrt{g_R^2 + g_1^2c^2}$, we see
\begin{align}
 & \frac{g_1c}{l}=s_{13}~. 
 \label{eq:triangle_eq}
\end{align}
Hence Eq.(\ref{eq:solution_theta23}) can be written as
\begin{align}
 & \tan{\theta _{23}} = - \frac{g_1 s}{l} + O(\delta ) = - \frac{s s_{13}}{c} + O(\delta )~, 
 \label{eq:tan_theta_23_2}
\end{align}
This can be rewritten as
\begin{align}
 & \frac{c^2}{c_{23}^2} =  1 - s^2 c_{13}^2 + O(\delta )~.
 \label{eq:solution_u_w}
\end{align}
In $\delta \sim 0$, then introducing new variables by
\begin{align}
 & s c_{13} = s'~, \quad \frac{c}{c_{23}} = c'
 \label{eq:def_theta'}
\end{align}
we see from Eq. (\ref{eq:solution_u_w})
\begin{align}
 & c'^2 + s'^2 = 1~,
 \label{eq:relation_c'_s'} 
\end{align}
Hence one can set $s', c'$ as $s'=\sin{\theta '}, c'=\cos{\theta '}$, respectively. Equation (\ref{eq:ratio_MZ_MW}) is now simply given by  
\begin{align}
 & \frac{M_Z}{M_W} = \frac{1}{c'}
 \label{eq:ratio_MZ_MW_2}
\end{align}
\indent Another useful relation is obtained from the low-energy $\nu e$ scattering amplitude with the $W$  boson exchange. This is given by, from (\ref{eq:Gauge_part}),
\begin{align}
 & \Big(\frac{g_L}{2\sqrt{2}} \cos{\gamma }\Big)^2 \frac{1}{M_W^2} = \frac{G}{\sqrt{2}}~,
  \label{eq:relation_G_gL} 
\end{align}
where $G$ is the Fermi constant, and $\gamma $ is the mixing angle between $W_L$ and $W_R$. \\
One can put $\cos{\gamma }\simeq 1$, since $\gamma $ is very small. Substituting $g_L=e_0/(sc_{13})=e_0/s'$ into 
Eq.(\ref{eq:relation_G_gL}) , we have
\begin{align}
 & M_W = 37.3\frac{1}{s'}~\mbox{GeV}~.
 \label{eq:value_MW}
\end{align}
%
\indent The coupling strength $g_{\nu Z}$ between $\nu $ and $Z$ can be seen from (\ref{eq:Gauge_part}) to be

\begin{align}
 & \bar{\nu }_L \gamma x_L \nu _L = g_{\nu Z} \bar{\nu }_L \gamma \nu _L Z~,
 \label{eq:coupling_strength_nu_Z} 
\end{align}
where
\begin{align}
 & g_{\nu Z} = \frac{1}{2}(-g_1 T_{12} + g_L T_{22}) = \frac{e_0}{2c's'}~.
 \label{eq:g_nu_Z} 
\end{align}
\indent The coupling strength between the electron and $Z$ is also calculated from (\ref{eq:Gauge_part}) as follows:
\begin{align}
 & \bar{e}_L \gamma y_L e_L + \bar{e}_R \gamma y_R e_R
 \label{eq:coupling_strength_e_Z} \\
 & = \bar{e}_L \gamma e_L Z \frac{1}{2} (g_1 T_{12} + g_L T_{22}) + \bar{e}_R \gamma e_R Z \frac{1}{2} (g_1 T_{12} + g_R T_{32})
 \nonumber \\
 & = (\alpha _L \bar{e}_L \gamma e_L + \alpha _R \bar{e}_R \gamma e_R ) Z~,
 \nonumber
\end{align}
with
\begin{align}
 & \alpha _L = \frac{e_0}{2c' s'}(c'^2 - s'^2)~,
 \label{eq:alpha} \\
 & \alpha _R = -\frac{e_0 s'}{c'}~,
 \nonumber
\end{align}
where use has been made of Eq.(\ref{eq:tan_theta_23_2}), i. e., $\tan{\theta _{23}}=-s s_{13}/c$.\\
Hence the $eZ$ coupling strengths $\alpha _{L,R}$ are also a function of $c'$ and $s'$. The same formula as $\alpha _{L,R}$ can be shown to hold for coupling strengths between the proton and $Z$ (see the Appendix). \\
\indent As far as light particles such as $W^{\pm}, Z, \nu _{L}$ and charged leptons (quarks) are 
concerned, all results are completely the same as those of the Weinberg-Salam theory with $SU(2)_L\times U(1)_Y$ in $\delta \sim 0$ if we take $s'=\sin{\theta _W}, c'=\cos{\theta _W}$, where $\theta _W$ is the Weinberg angle. Here WS gauge coupling constants $g$ and $g'$ are given by
\begin{align}
 & g' \equiv g_1c_{13}c_{23}=\frac{e_{0}}{c}c_{23} = \frac{e_0}{c'}=\frac{e_{0}}{\cos{\theta _W}}=\frac{e_{0}}{0.88}=0.34~,
 \label{eq:def_g'} \\
 & g=\frac{e_0}{\sin{\theta _W}}=\frac{e_{0}}{sc_{13}}=g_L=\frac{e_{0}}{0.48}=0.63~,
 \nonumber
\end{align}
respectively, around mass scales of weak gauge bosons $M_W, M_Z$. Both equations $g=g_L$ and $g'=g_1c_{13}c_{23}$ are our new results together with $\theta '=\theta _W$. \\
\indent Here we can show that any mixing angle is expressed in terms of $\theta _W$ and $\varepsilon \equiv g_L/g_R$. At first we have
\begin{align}
 & \varepsilon = \frac{g_L}{g_R} = \frac{s_{13}}{sc_{13}}=\frac{s_{13}}{s'}=\frac{s_{13}}{s_W}~,
 \label{eq:ratio_epsilon_s_W}
\end{align}
so that
\begin{align}
 & s_{13} = \varepsilon s_W~.
 \label{eq:cal_c{13}}
\end{align}
Secondary, from Eq.(\ref{eq:ratio_epsilon_s_W}) we get
\begin{align}
 & \varepsilon ^2 = \frac{1}{s_W^2} - \frac{1}{s^2}~,
 \label{eq:deffence_expsion_epsilon}
\end{align}
to follow
\begin{align}
 & s = s_{12} = \frac{s_W}{\sqrt{1-\varepsilon ^2 s_W^2}}~.
 \label{eq:cal_s12_s_w}
\end{align}
Third, from the definition (\ref{eq:def_theta'}) we have
\begin{align}
 & c_{23}^2 = \frac{c^2}{c'^2} = \frac{1 - s^2}{c'^2}~.
 \label{eq:cal_c23_s_c'}
\end{align}
Substituting Eq.(\ref{eq:cal_s12_s_w}) into the above, we get
\begin{align}
 & c_{23} = \sqrt{\frac{1 - \varepsilon ^2 \tan^2{\theta _W}}{1 - \varepsilon ^2 \sin^2{\theta _W}}}~.
 \label{eq:cal_c23_epsilon_sita}
\end{align}
\indent To sum up we have
\begin{align}
 & \theta ' = \theta _{W}~, \quad \varepsilon  = \frac{g_L}{g_R}~,
 \label{eq:three_mixing_angles} \\
 & \theta _{12} = \sin^{-1}{\Big(\frac{\sin{\theta _W}}{\sqrt{1 - \varepsilon ^2 \sin^2{\theta _W}}}\Big)}~,
 \nonumber \\
 & \theta _{13} = \sin^{-1}{\big(\varepsilon \sin{\theta _W}\big)}~,
 \nonumber \\
 & \theta _{23} = \cos^{-1}{\sqrt{\frac{1 - \varepsilon ^2 \tan^2{\theta _W}}{1 - \varepsilon ^2 \sin^2{\theta _W}}}}
 \nonumber
\end{align}
\indent Finally let us discuss about the ratio of gauge coupling strengths $\varepsilon = g_L/g_R$.  From Eq.(\ref{eq:deffence_expsion_epsilon}) we see that it cannot be fixed by $\theta _W$ only. However, the $s$ can be expressed in terms of a mass ratio for heavy gauge bosons
\begin{align}
 & \frac{M_{W'}} {M_{Z'}}= \frac{c_{13} c_{23}}{\sqrt{2}}~,
 \label{eq:ratio_mass_MZ'_MW'}
\end{align}
which comes from Eqs.(\ref{eq:mass_MZ'_3}) and (\ref{eq:mass_M_W'}). By definitions (\ref{eq:def_theta'}) we have  
\begin{align}
 & c_{13}c_{23} = \frac{s'c}{sc'} = \frac{\tan{\theta '}}{\tan{\theta }}~, 
 \label{eq:relation_c13_c23}
\end{align}
so that
\begin{align}
 & \varepsilon ^2 = \frac{1}{s'^2} - \frac{1}{s^2} = \cot^2{\theta '} -\cot^2{\theta} = \cot^2{\theta '}\Big[ 1- 2\frac{M_{W'}^2}{M_{Z'}^2}\Big]~.
 \label{eq:relation_epsilon2} 
\end{align}
Hence we get
\begin{align}
 & \frac{M_{W'}^2}{M_{Z'}^2} = \frac{1}{2} \big( 1 - \varepsilon ^2 \tan^2{\theta '} \big)~.
 \label{eq:ratio_mass2}
\end{align}
We know the experimental value of $\theta '$ to be $\cot{\theta '}=1.83$ ($\sin^2{\theta '}=0.23$). Then, Duka et al.\cite{ref:Duka} worked with $\epsilon =1$ to obtain $x=M_{W'}/M_{Z'}=0.59$. Deppisch et al.\cite{ref:Deppisch} found the pair $(x, \varepsilon )=(0.30, 1.66)$ based on the $SO$(10)  unified model, and Dev et al.\cite{ref:Dev} found $(x, \varepsilon )=(0.63, 1.23)$. On the other hand Patra et al.\cite{ref:Patra} worked in a region $M_{W'}> M_{Z'}$, by considering another model where Eq.(\ref{eq:ratio_mass2}) does not hold. \\
%
%
\section{Concluding remarks} 
%
\indent Our new results are summarized as follows: \\
 \\
1.	The mass matrix for neutral gauge bosons becomes diagonal if
\begin{align}
 & \tan{\theta _{23}} = - \frac{s_{12} s_{13}}{c_{12}} + O(\delta )~, \quad \delta = \frac{\kappa _1^2 + \kappa _2^2}{4V_R^2}=5.8\times 10^{-4}<<1 
 \nonumber
\end{align}
Under this constraint we have defined uniquely a new mixing angle $\theta '$ by
\begin{align}
 & s' = s c_{13}~, \quad c' = \frac{c}{c_{23}}, \quad s'^2 + c'^2 = 1~.
 \nonumber
\end{align}
Then the light gauge boson world is described by $\theta '$ only as below
\begin{align}
 & M_W = \frac{37.3}{s'}~\mbox{GeV}~,
 \nonumber \\
 & \frac{M_Z}{M_W} = \frac{1}{c'}~,
 \nonumber \\
 & g_{\nu Z} = - \frac{e_0}{2c's'}~,
 \nonumber \\
 & g(pZ)_L = \frac{e_{0}(c'^2 - s'^2)}{2s' c'}~, \quad g(pZ)_R = - \frac{e_0 s'}{c'}~,
 \nonumber
\end{align}
All results are completely the same as those of the Weinberg-Salam theory with $SU(2)_L\times U(1)_Y$, if we take $s'\equiv \sin{\theta _W}$, $c'=\cos{\theta _W}$ where $\theta _W$ is the Weinberg angle. \\
The WS gauge coupling constants $g$ and $g'$ are given by
\begin{align}
 & g' \equiv g_1c_{13}c_{23}=\frac{e_{0}}{c_{12}}c_{23} = \frac{e_0}{c'}=\frac{e_{0}}{\cos{\theta _W}}=0.34~,
 \nonumber \\
 & g=\frac{e_0}{\sin{\theta _W}}=\frac{e_{0}}{sc_{13}}=g_L=\frac{e_{0}}{0.48}=0.63~,
 \nonumber
\end{align}
\\
2.	Any mixing angle is expressed in terms of $\theta _W$ and $\varepsilon =g_L/g_R$ as below
\begin{align}
 & \theta ' = \theta _{W}~, 
 \nonumber \\
 & \theta _{12} = \sin^{-1}{\Big(\frac{\sin{\theta _W}}{\sqrt{1 - \varepsilon ^2 \sin^2{\theta _W}}}\Big)}~,
 \nonumber \\
 & \theta _{13} = \sin^{-1}{\big(\varepsilon \sin{\theta _W}\big)}~,
 \nonumber \\
 & \theta _{23} = \cos^{-1}{\sqrt{\frac{1 - \varepsilon ^2 \tan^2{\theta _W}}{1 - \varepsilon ^2 \sin^2{\theta _W}}}}
 \nonumber
\end{align}
\\
3. We have given the mass formula for heavy gauge bosons
\begin{align}
 & \frac{M_{W'}^2}{M_{Z'}^2} = \frac{1}{2} \big( 1 - \varepsilon ^2 \tan^2{\theta '} \big)~.
 \nonumber
\end{align}
A similar formula like this has been so far derived by many authors in different ways from ours. Here we have given its exact proof based on our formula $\tan{\theta '}=s_{12} s_{13} c_{23}/c_{12}$. \\ \\
4. Mass formulas (\ref{eq:mass_Z_2}) and (\ref{eq:mass_MZ'_3}) for $M_Z$ and $M_{Z'}$ are new forms expressed in terms of mixing angles. \\
 \\
5. Our general observations are that the world of light gauge bosons is described by $\theta '$ only, whereas the world of heavy gauge bosons is described by two parameters  $\varepsilon $ and $\theta '$, together with $V_R$. The first notice of the $Z'$ mass can be seen in Ref. \cite{ref:Altarelli} with the $\varepsilon =1$ model, and in Ref.\cite{ref:Barger} with the extra $U(1)$ model. \\
 \\
\indent The LRSM can be constructed from the geometric point of view of the gauge theory in  $M_4\times Z_2\times Z_2$, where $M_4$ is the four-dimensional Minkowski space and $Z_2\times Z_2$ is the discrete space with four points\cite{ref:Konisi}. The three Higgs fields $\phi , \Delta _L$ and $\Delta _R$ can be regarded as gauge fields in $Z_2\times Z_2$ . The Higgs potential, therefore, should be of the Yang-Mills type, which contains eleven free parameters. This should be compared with the general Higgs potential \cite{ref:Deshpande}, which contains eighteen parameters.
\section*{Acknowledgments}

We would like to express our deep gratitude to T. Okamura for many valuable discussions.

\appendix
\section{Electromagnetic couplings with fermions and the proton-$Z$ coupling strengths}\label{sec:appendA}
%
The fermionic Lagrangian with gauge couplings is given by
\begin{align}
 & L_F = i \bar{\psi}_L \gamma ^\mu \Big(\partial _\mu - i\frac{1}{2} g_1 Y_F B_\mu - i\frac{1}{2}  g_L \tau _\alpha {W_\mu ^\alpha }_L \Big) \psi_L
 \label{eq:Action_fermion_part2} \\
& \quad  + i \bar{\psi}_R \gamma ^\mu \Big(\partial _\mu - i\frac{1}{2} g_1 Y_F B_\mu - i\frac{1}{2}  g_R \tau _\alpha{W_\mu ^\alpha  }_R \Big) \psi _R~,
 \quad \psi =
 	\left(
   \begin{array}{c}
      u \\
      d \\
   \end{array}
  \right)~,
   	\left(
   \begin{array}{c}
      \nu  \\
      e \\
   \end{array}
  \right)~,
   	\left(
   \begin{array}{c}
      p \\
      n \\
   \end{array}
  \right)~, \cdots
 \nonumber  
\end{align}
Gauge coupling terms $X$ are included in the form for the lepton case
\begin{align}
 & X = \bar{\nu }_L \gamma x_L \nu_L + \bar{e}_L \gamma y_L e_L + \frac{g_L}{\sqrt{2}}\bar{\nu }_L \gamma  W_L^+ e_L + \frac{g_L}{\sqrt{2}}\bar{e}_L \gamma W_L^- \nu _L 
 \label{eq:Gauge_part} \\
 & \quad + \bar{\nu }_R \gamma x_R \nu_R + \bar{e}_R \gamma y_R e_R + \frac{g_R}{\sqrt{2}}\bar{\nu }_R \gamma W_R^+ e_R + \frac{g_R}{\sqrt{2}}\bar{e}_R \gamma W_R^- \nu _R ~,
 \nonumber  
\end{align}
where $W_{L,R}^{\pm} = (W^1 \mp i W^2)_{L,R}/\sqrt{2}$, and
\begin{align}
 & x_L = \frac{1}{2}(g_1 Y_F B + g_L W_L^3) = \frac{1}{2}g_1 Y_F (T_{11}A + T_{12}Z + T_{13} Z') + \frac{1}{2}g_L  (T_{21}A + T_{22}Z + T_{23} Z')
 \label{eq:def_xL} \\
 & \quad = \frac{1}{2}A(g_1 Y_F T_{11} + g_L T_{21}) + \frac{1}{2}Z (g_1 Y_F T_{12} + g_L T_{22}) + \frac{1}{2}Z' (g_1 Y_F T_{13} + g_L T_{23})~,
  \nonumber \\
& y_L = \frac{1}{2}(g_1 Y_F B - g_L W_L^3)  = \frac{1}{2}g_1 Y_F (T_{11}A + T_{12}Z + T_{13} Z') - \frac{1}{2}g_L  (T_{21}A + T_{22}Z + T_{23} Z')
 \label{eq:def_yL} \\
 & \quad = \frac{1}{2}A(g_1 Y_F T_{11} - g_L T_{21}) + \frac{1}{2}Z (g_1 Y_F T_{12} - g_L T_{22}) + \frac{1}{2}Z' (g_1 Y_F T_{13} - g_L T_{23})~,
 \nonumber \\
 & x_R = \frac{1}{2}(g_1 Y_F B + g_R W_R^3)  = \frac{1}{2}g_1 Y_F (T_{11}A + T_{12}Z + T_{13} Z') + \frac{1}{2}g_L  (T_{31}A + T_{32}Z + T_{33} Z')
 \label{eq:def_xR} \\
 & \quad = \frac{1}{2}A(g_1 Y_F T_{11} + g_R T_{31}) + \frac{1}{2}Z (g_1 Y_F T_{12} + g_R T_{32}) + \frac{1}{2}Z' (g_1 Y_F T_{13} + g_R T_{33})~,
 \nonumber \\
& y_R = \frac{1}{2}(g_1 Y_F B - g_R W_R^3) = \frac{1}{2}g_1 Y_F (T_{11}A + T_{12}Z + T_{13} Z') - \frac{1}{2}g_L  (T_{31}A + T_{32}Z + T_{33} Z')
 \label{eq:def_yR} \\
& \quad = \frac{1}{2}A(g_1 Y_F T_{11} - g_R T_{31}) + \frac{1}{2}Z (g_1 Y_F T_{12} - g_R T_{32}) + \frac{1}{2}Z' (g_1 Y_F T_{13} - g_R T_{33})~,
 \nonumber
\end{align}
with $Y_F=-1$. Collecting the electromagnetic terms and substituting them into (\ref{eq:Gauge_part}), we have constraint equations

\begin{align}
 & \frac{1}{2}(g_1 Y_F T_{11} + g_L T_{21} ) = \frac{1}{2}( g_1 Y_F T_{11} + g_R T_{31} ) = 0~,
 \label{eq:constrain1} \\
 & \frac{1}{2}(g_1 Y_F T_{11} - g_L T_{21} ) = \frac{1}{2}( g_1 Y_F T_{11} - g_R T_{31} )= - e_0~,
 \label{eq:constrain2} 
\end{align}
Since the lepton takes $Y_F=-1$, we get
\begin{align}
 & g_1 T_{11} = g_L T_{21} = g_R T_{31} = e_0~.
 \label{eq:constrain3}
\end{align}
\indent The coupling strength between the proton and the $Z$ boson can be calculated in such a way that $\psi =(p \ n)_{L,R}$ are $SU(2)_L\times SU(2)_R$ doublets with $Y_F=+1$. Neutral gauge coupling terms $X$ are now included in 
\begin{align}
 & X = \bar{p}_L \gamma x_L p_L + \bar{p}_R \gamma x_R p_R~,
 \label{eq:Neutral_gauge_X}
\end{align}
where
\begin{align}
 & x_L = Z\frac{1}{2}(g_1 T_{12} + g_L T_{22}) =  Z \frac{e_0}{2c's'}(c'^2 - s'^2) \equiv Z \alpha _L~,
 \label{eq:cal_xL}
\end{align}
is the same calculation as Eq. (6.10), while the second part goes as follows:
\begin{align}
 & x_R = Z\frac{1}{2}(g_1 T_{12} + g_R T_{32}) = Z \frac{e_0}{2}(\frac{1}{c c_{13}}(-sc_{13} -cs_{13}s_{23}) + \frac{1}{s_{13}}s_{23}c_{13})
 \label{eq:cal_xR} \\
 & =Z \frac{e_0 c_{23}}{2}\Big[ -\frac{s}{c c_{13}} + \big(\frac{c_{13}}{s_{13}}-\frac{s_{13}}{c_{13}}\big) \tan{\theta _{23}}\Big]~.
 \nonumber
\end{align}
Substituting $\tan{\theta _{23}}= -s s_{13}/c$ into above, we have
\begin{align}
 &  x_R = Z \frac{e_0 c_{23}}{2}\Big[ -\frac{s}{c c_{13}} - \big(\frac{c_{13}}{s_{13}}-\frac{s_{13}}{c_{13}}\big) \frac{s s_{13}}{c}\Big] = -Z e_0 \frac{s'}{c'} \equiv Z \alpha _R ~,
 \label{eq:cal_xR2}
\end{align}
\indent To sum up, neutral current strengths between the proton and the $Z$ boson are given by
\begin{align}
 & \alpha _L = \frac{e_0}{2c' s'}(c'^2 - s'^2)~,
 \label{eq:neutral_current_strength} \\
 & \alpha _R = -e_0 \frac{s'}{c'}
 \nonumber
\end{align}
These are completely the same forms as Eq.(\ref{eq:alpha}) for the electron and $Z$.
%
%

\end{document}